\documentclass[12pt]{article}
\usepackage{amssymb}
\begin{document}
\title{\bf Conserved Quantities in $f(R)$ Gravity via Noether Symmetry}
\author{M. Farasat
Shamir\thanks{farasat.shamir@nu.edu.pk}, Adil Jhangeer
\thanks{adil.jahangeer@nu.edu.pk} and Akhlaq Ahmad Bhatti \thanks{akhlaq.ahmad@nu.edu.pk} \\\\  Department of
Sciences \& Humanities, \\National University of Computer \&
Emerging Sciences,\\ Lahore Campus, Pakistan.\\ Tel:
92-42-111-128-128(Ext.229), Fax: 92-42-35165232.}
\date{}
\maketitle
\begin{abstract}
This paper is devoted to investigate $f(R)$ gravity using Noether
symmetry approach. For this purpose, we consider Friedmann
Robertson-Walker (FRW) universe and spherically symmetric
spacetimes. The Noether symmetry generators are evaluated for some
specific choice of  $f(R)$ models in the presence of gauge term.
Further, we calculate the corresponding conserved quantities in each
case. Moreover, the importance and stability criteria of these
models are discussed.
\end{abstract}


\section{Introduction}

Astrophysical data from different sources such as cosmic microwave
background fluctuations \cite{2}, Supernovae Ia (SNIa) \cite{sn}
experiments, X-ray experiments \cite{xray} and large scale structure
\cite{3} indicate that our universe is currently expanding with an
accelerated rate. Higher dimensional theories \cite{highdim} like
M-theory or string theory may explain this accelerated expansion.
Another explanation comes from modification of Einstein's theory
with some inverse curvature terms which cause increase in gravity
\cite{invR}. However, modified gravity with inverse curvature terms
is known to be unstable and do not pass some solar system tests
\cite{unstable}. This discrepancy can be removed by including higher
derivative terms. In particular, the viability \cite{viable} can be
achieved with squared curvature terms. It is now thought that the
current cosmic expansion can be justified if some suitable powers of
curvature are added to the usual Einstein-Hilbert action \cite{5}.
Thus it would be interesting to investigate the universe in the
context of modified or alternative theories of gravity. The $f(R)$
theory of gravity, which involves a generic function of Ricci scalar
in standard Einstein-Hilbert lagrangian, is an attractive choice.

In recent years, many authors investigated $f(R)$ gravity in
different contexts. Felice and Tsujikawa \cite{Felice and Tsujikawa}
gave a detailed review about $f(R)$ theories of gravity. A similar
work has been reported by Sotiriou and Faraoni \cite{Sotiriou and
Faraoni}. Hendi and Momeni \cite{M} explored black hole solutions in
$f(R)$ gravity. Moon and Myung \cite{moon} gave the stability
analysis of the Schwarzschild black hole in this theory. Geodesic
deviation equation in metric $f(R)$ gravity is obtained by Guarnizo
et al. \cite{GD}. Upadhye and Hu \cite{stars} investigated the
existence of relativistic stars in $f(R)$ gravity. The metric $f(R)$
theories of gravity are generalized to five dimensional spacetimes
by Huang et al. \cite{fiveD}. They showed that expansion and
contraction of the extra dimension prescribed a smooth transition
from deceleration phase to acceleration phase. The stability
conditions for $f(R)$ models have been discussed by Starobinsky
\cite{Starobinsky}. Multam\"{a}ki and Vilja \cite{9, 10} explored
spherically symmetric vacuum and non-vacuum solutions in $f(R)$
theory of gravity. Shojai and Shojai \cite{spherically} calculated
exact spherically symmetric interior solutions in metric version of
$f(R)$ gravitational theory. Azadi et al. \cite{13} investigated
cylindrically symmetric vacuum solutions in this theory. Plane
symmetric solutions are studied by Sharif and Shamir \cite{15}. The
same authors \cite{14, 18} investigated the solutions for Bianchi
types $I$ and $V$ models for both vacuum and non-vacuum case.

The field equations in $f(R)$ gravity are fourth order partial
differential equations (PDEs) when the function is assumed to have
the terms like $R^2$. However, the order could be higher if the
terms like $R^3, R^4$ etc. are included. On the other hand, Lie's
theory gives a systematic and mathematical way to investigate the
solutions of differential equations. The application of Lie group
theory for the solution of nonlinear ordinary differential equation
is one of the most fascinating and significant area of research. It
is mentioned here that from Lie's theory one can not only construct
a class of exact solutions but can also find new solutions using
different invariant transformations. It also gives most widely
applicable technique to find the closed form solution of
differential equations. Investigation of these solutions plays a
vital role for the understanding of the physical aspects of these
differential equations.

Noether symmetry approach is an important aspect of Lie theory. This
is the most elegant and systematic approach to compute conserved
vectors, given by Noether in $1918$. The conservation laws play a
vital role in the study of physical phenomenon. The integrability
for PDEs depends on number of conservation laws. Another important
aspect of conservation laws is that they are helpful in the
numerical integration of PDEs. There are number of methods developed
for the construction of conservation laws such as Noether theorem
\cite{UD,IV} for variational problems, partial Noether theorem for
variational and non variational structures \cite{NC} and multiplier
approach \cite{multiplier}. Computer packages for the construction
of conserved quantities are reported by many authors. Some of them
are: Wolf \cite{CC1}, Wolf et al. \cite{CC2}, G\"{o}ktas et al.
\cite{Goktas} and Hereman et al. \cite{Hereman1, Hereman2,
Hereman3}. The Maple code to compute conservation laws based on
multipliers approach was introduced by Cheviakov \cite{Cheviakov}.

Noether theorem states that any differentiable symmetry of the
action of a physical system has a corresponding conservation law.
The main feature of this theorem is that it may provide information
regarding the conservation laws in theory of relativity.
Conservation laws of linear and angular momentum can be well
explained by the translational and rotational symmetries using
Noether theorem \cite{Noethertheorem1}. It has many applications in
theoretical physics. In recent years, many authors have used this
theorem in different cosmological contexts. Capozziello et al.
\cite{20} discussed $f(R)$ gravity for spherically symmetric
spacetime using Noether symmetry. Flat FRW universe has been
discussed in Palatini $f(R)$ gravity by Kucukakca and Camci
\cite{platini}. Jamil et al. \cite{21} investigated $f(R)$ Tachyon
model via Noether symmetry approach. Hussain et al. \cite{22} found
Noether symmetries for flat FRW model using gauge term in metric
$f(R)$ gravity. Energy distribution of Bardeen model is given by
Sharif and Waheed \cite{23} using approximate symmetry method. The
same authors \cite{24} re-scaled the energy in the stringy charged
black hole solutions using approximate symmetries. In a recent
paper, we \cite{farasat} have found a new class of plane symmetric
solutions in metric $f(R)$ gravity using Lie point symmetries.

In this paper, we focus our attention to investigate the Noether
symmetries of FRW and spherically symmetric spacetimes in the
context of metric $f(R)$ gravity. The paper is organized as follows:
In Section \textbf{2}, we present some basics of $f(R)$ theory of
gravity. Sections \textbf{3} and \textbf{4} are used to calculate
Noether symmetries of FRW and spherically symmetric spacetimes
respectively. In the last section, we summarize the results.

\section{$f(R)$ Gravity and Field Equations}

The action for four dimensional $f(R)$ theory of gravity in
gravitational units $(8\pi G=1)$ is given by \cite{20}
\begin{equation}\label{1}
S_{f(R)}=\int\sqrt{-g}(f(R)+L_{m})d^4x,
\end{equation}
where $L_{m}$ is the matter Lagrangian and $f(R)$ is a general
function of the Ricci scalar. The standard Einstein-Hilbert action
can be obtained by taking $f(R)=R$. Variation of this action with
respect to the metric tensor yields the field equations
\begin{equation}\label{2}
f'(R)R_{\mu\nu}-\frac{1}{2}f(R)g_{\mu\nu}-\nabla_{\mu}
\nabla_{\nu}f'(R)+g_{\mu\nu}\Box f'(R)=\kappa T^m_{\mu\nu},
\end{equation}
where prime denotes derivative with respect to $R$, $\kappa$ is a
coupling constant in gravitational units, $T^m_{\mu\nu}$ is the
standard standard matter energy-momentum tensor and
\begin{equation}\label{3}
\quad\Box\equiv\nabla^{\mu}\nabla_{\mu}
\end{equation}
with $\nabla_{\mu}$ defined as the covariant derivative. The field
equations can be expressed in an alternative form familiar with
General Relativity (GR) field equations as
\begin{equation}\label{4}
G_{\mu\nu}=R_{\mu\nu}-\frac{1}{2}g_{\mu\nu}R=T^c_{\mu\nu}+\tilde{T}^m_{\mu\nu},
\end{equation}
where $\tilde{T}^m_{\mu\nu}=T^m_{\mu\nu}/f'(R)$ and
energy-momentum tensor for gravitational fluid is given by
\begin{equation}\label{5}
T^c_{\mu\nu}=\frac{1}{f'(R)}\bigg[\frac{1}{2}g_{\mu\nu}\bigg(f(R)-Rf'(R)\bigg)+
f'(R)^{;\alpha\beta}\bigg(g_{\alpha\mu}g_{\beta\nu}-g_{\mu\nu}g_{\alpha\beta}\bigg)\bigg].
\end{equation}
It is clear from Eq.(\ref{4}) that energy-momentum tensor for
gravitational fluid $T^c_{\mu\nu}$ contributes matter part from
geometric origin. This approach seems interesting as it may provide
all the matter components which are required to investigate the dark
part of our universe. Thus it is expected that $f(R)$ theory of
gravity may give fruitful results to understand the phenomenon of
expansion of universe.

\section{Noether FRW Symmetries}

In this section, we shall find the Noether symmetries of FRW
spacetime. The FRW metric is given by
\begin{equation}\label{33}
ds^{2}=dt^{2}-a^2\bigg[\frac{dr^{2}}{1-kr^2}+r^{2}d\Omega^{2}\bigg],
\end{equation}
where $a$ is function of cosmic time $t$ and known as scale factor
of universe and $d\Omega^{2}=r^2(d\theta^2+\sin^2\theta d\phi^2)$.
The curvature parameter $k$ is $0$, $1$ or $-1$, which represents
flat, open or closed universe respectively. The corresponding
Lagrangian is given by \cite{quintessence}
\begin{equation}\label{l34}
L=6a\dot{a}^2f'+6a^2\dot{R}\dot{a}f''+a^3(f-Rf')-6kaf'+ a^3P.
\end{equation}
Here dot denotes derivative with respect to $t$ and the fluid
pressure $P$ is given by
\begin{equation} \label{37}
P=m\omega a^{-3(1+\omega)},
\end{equation}
where $m$ is an arbitrary real constant while $\omega$ is the
equation of state parameter. Noether symmetry generator of
Eq.(\ref{l34}) is given by
\begin{equation}\label{(2T)}
X=\tau(t, a, R)\frac{\partial}{\partial t}+ \psi(t, a,
R)\frac{\partial}{\partial a}+\phi(t, a, R)\frac{\partial}{\partial
R}\cdot
\end{equation}
We can find Noether symmetries by using the following equation:
\begin{equation}\label{(l22T)}
X^{[1]}L + (D\tau)L=DB(t, a, R).
\end{equation}
where $X^{[1]}$ is the first prolongation \cite{Olver} given by
\begin{equation}\label{(2922T)}
X^{[1]}=X+\dot{\psi}(t, a, R)\frac{\partial}{\partial
\dot{a}}+\dot{\phi}(t, a, R)\frac{\partial}{\partial \dot{R}},
\end{equation}
where
\begin{eqnarray}
\dot{\psi}=\frac{\partial \psi}{\partial
t}+\dot{a}\frac{\partial\psi}{\partial
a}-\dot{a}\frac{\partial\tau}{\partial
t}+\dot{R}\frac{\partial\psi}{\partial
R}-{\dot{a}}^2\frac{\partial\tau}{\partial a}-
\dot{a}\dot{R}\frac{\partial\tau}{\partial R},\\
\dot{\phi}=\frac{\partial\phi}{\partial
t}+\dot{a}\frac{\partial\phi}{\partial
a}+\dot{R}\frac{\partial\phi}{\partial
R}-\dot{R}\frac{\partial\tau}{\partial
t}-{\dot{R}}^2\frac{\partial\tau}{\partial
R}-\dot{a}\dot{R}\frac{\partial\tau}{\partial a}.
\end{eqnarray}
$B$ is called the gauge function with $D$ defined as
\begin{equation}
D\equiv \frac{\partial}{\partial
t}+\dot{a}\frac{\partial}{\partial
a}+\dot{R}\frac{\partial}{\partial R}.
\end{equation}
The first integral is also known as conserved quantity associated
with $X$ and is defined as
\begin{equation}\label{(integral)}
I=B-\tau L -(\psi-\tau\dot{a})\frac{\partial L}{\partial
\dot{a}}-(\phi-\tau\dot{R})\frac{\partial L}{\partial \dot{R}}.
\end{equation}

Using ($\ref{l34}$) in Eq.($\ref{(l22T)}$), we obtain an over
determined system of linear PDEs, i.e.
\begin{eqnarray}
\label{(D1)}\tau_a&=&0,\\\label{(D2)}\tau_R&=&0,\\\label{(D3)}f''\psi_R&=&0,
\\\label{(D4)}6a^2f''\psi_t&=&B_R,\\
\label{(D5)}12af'\psi_t+6a^2f''\phi_t&=&B_a,\\\label{(D6)} \psi
f'+\phi
af''+2af'\psi_{a}-af'\tau_{t}+a^2f''\phi_a&=&0,\\\label{(D7)}
2af''\psi+a^2f'''\phi+a^2f''(\psi_a+\phi_R-\tau_{t})+2af'\psi_R&=&0,\\\nonumber
a^2(3\psi+a\tau_t)(f-Rf')-a^3Rf''\phi+\tau_t(-6kaf'+\omega
ma^{-3\omega})-\\\psi(6kf'+3\omega^2ma^{-3\omega-1})&=&B_t\label{(D8)}.
\end{eqnarray}
We solve these equations using different assumptions. From
Eq.($\ref{(D3)}$), we have $f''=0$ with $\psi_R\neq0$. This further
gives $f'=0$ by using Eq.($\ref{(D7)}$). Solving determining
equations with these conditions, we obtain trivial solution, i.e.
\begin{eqnarray}\label{(29T)}
\tau=0,~~\psi=0, ~~\phi=0,
\end{eqnarray}
with zero gauge term. Hence we investigate the solution of
determining equations by taking $\psi_R=0$ with $f''\neq0$. The
solutions are discussed mainly for two different cases, namely flat
vacuum universe and non-flat non-vacuum universe.
\\\\
\textbf{Case 1: Flat Vacuum Universe ($k=0=\omega$)}
\\\\
Here we solve the determining equations for flat vacuum case. It
is mentioned here that this case has already been discussed by
Jamil et. al \cite{22} but obtained results suggest that the gauge
term is zero. However, we have explored a more general solution of
the determining equations for $f(R)=f_0R^{3/2}$ which gives a
non-zero gauge term. The solution in this case is given by
\begin{eqnarray}\label{(Sol)}
\tau&=&c_1t+c_2,\\\psi&=&\frac{2c_1a^2+3c_3t+3c_4}{3a},\\\phi&=&-2R\frac{c_1a^2+c_3t+c_4}{a^2},\\B&=&9c_3a\sqrt{R}+c_5.
\end{eqnarray}
The Noether symmetry generators turn out to be
\begin{eqnarray}\label{(NGg)}
X_1&=&t\frac{\partial}{\partial
t}+\frac{2}{3}a\frac{\partial}{\partial
a}-2R\frac{\partial}{\partial R},\\X_2&=&\frac{\partial}{\partial
t},\\X_3&=&ta^{-1}\frac{\partial}{\partial
a}-2Rta^{-2}\frac{\partial}{\partial
R},\\X_4&=&a^{-1}\frac{\partial}{\partial
a}-2Ra^{-2}\frac{\partial}{\partial R}.
\end{eqnarray}
These generators form a four dimensional algebra with the
following commutator table:
\begin{center}
\begin{tabular}{|l||l|l|l|l|l|}
 \hline
\multicolumn{5}{|c|}{ Table $1$: Commutator Table } \\
\hline  ~~& $X_{1}$ & $X_{2}$ & $X_{3}$& $X_{4}$
 \\ \hline \hline
 {$X_{1}$}  ~~ & 0 & $X_2$ & $\frac{X_3}{3}$ & $\frac{4X_4}{3}$
 \\ \hline

 {$X_{2}$}  ~~ & $-X_2$ & 0 & $-X_4$ & 0
 \\ \hline

 {$X_{3}$}  ~~ & $-\frac{X_3}{3}$ & $X_4$ & 0  & 0
 \\ \hline

 {$X_{4}$}  ~~ & $\frac{-4X_4}{3}$ & 0 & 0 & 0
 \\ \hline

 \end{tabular}
\end{center}
\vspace{2.5 mm} where Lie bracket $[X_{i},~X_{j}]$ is defined by the
following unique relation
\begin{eqnarray}
[X_{i},~X_{j}]=X_{j}\bigg(X_{i}\bigg)-X_{i}\bigg(X_{j}\bigg),
\nonumber
\end{eqnarray}
where $i,j=1,~2,~3,4.$ Moreover, the first integrals in this case
are
\begin{eqnarray}\nonumber
I_{1}&=&9at{\dot{a}}^2R^{\frac{1}{2}}+\frac{1}{2}a^3tR^{\frac{3}{2}}+
\frac{9}{2}a^2t\dot{a}\dot{R}R^{\frac{-1}{2}}-3a^2\dot{a}R^{\frac{1}{2}}-3a^3\dot{R}R^{\frac{-1}{2}},\\\nonumber
I_{2}&=&-9a{\dot{a}}^2R^{\frac{1}{2}}+\frac{1}{2}a^3tR^{\frac{3}{2}}-
\frac{9}{2}a^2\dot{a}\dot{R}R^{\frac{-1}{2}},\\\nonumber
I_{3}&=&9aR^{\frac{1}{2}}-9t\dot{a}R^{\frac{1}{2}}-
\frac{9}{2}ta\dot{R}R^{\frac{-1}{2}},\\\nonumber
I_{4}&=&-9\dot{a}R^{\frac{1}{2}}-
\frac{9}{2}a\dot{R}R^{\frac{-1}{2}}.
\end{eqnarray}
\textbf{Case 2: Non-Flat Non-Vacuum Universe
($k\neq0,~~\omega\neq0$)}\\\\ For this case, the determining
equations yield a solution
\begin{eqnarray}\label{(Sol1)}
\tau&=&c_1,~~~~~~~~B=c_2,\\\psi&=&0,~~~~~~~~~\phi=0.
\end{eqnarray}
Here the gauge term turns out to be constant which can be taken
zero. It is mentioned here that this solution is for an arbitrary
$f(R)$. The Noether symmetry generator turns out to be
\begin{equation}
X=\frac{\partial}{\partial t}.
\end{equation}
The corresponding first integral becomes
\begin{equation}
I=6a{\dot{a}}^2f'+6a^2{\dot{a}}\dot{R}f''-a^3(f-Rf')+6kaf'-m\omega
a^{-3\omega}.
\end{equation}

\section{Noether Symmetries of Spherically Symmetric Spacetime}

In this section, we shall find the Noether symmetries of static
spherically symmetric spacetime \cite{Zerbini}
\begin{equation}\label{z33}
ds^{2}=Adt^{2}-\bigg[\frac{dr^{2}}{A}+r^{2}d\Omega^{2}\bigg],
\end{equation}
where $d\Omega^{2}=r^2(d\theta^2+\sin^2\theta d\phi)$ and $A$ is the
function of $r$. The corresponding Lagrangian is given by
\begin{equation}\label{z34}
L=r^2(f-Rf')+2f'\bigg(1-A-r\frac{dA}{dr}\bigg)+
f''r^2\bigg(\frac{dR}{dr}\bigg)\bigg(\frac{dA}{dr}\bigg),
\end{equation}
here prime denotes derivative with respect to $R$.\\\\ Corresponding
Noether symmetry generator is given by
\begin{equation}\label{(z2T)}
X=\tau(r, R, A)\frac{\partial}{\partial r}+ \psi(r, R,
A)\frac{\partial}{\partial R}+\phi(r, R, A)\frac{\partial}{\partial
A}.
\end{equation}
The Noether symmetries can be computed by using the following
equation
\begin{equation}\label{(22T)}
X^{[1]}L + (D\tau)L=DB(t, a, R),
\end{equation}
where $X^{[1]}$ is the first prolongation \cite{Olver} given by
\begin{equation}\label{(22lpT)}
X^{[1]}=X+\acute{\psi}(r, R, A)\frac{\partial}{\partial
\acute{R}}+\acute{\phi}(r, R, A)\frac{\partial}{\partial \dot{A}},
\end{equation}
in which $ \acute{} $ represents derivative with respect to $r$ and
\begin{eqnarray}
\acute{\psi}=\frac{\partial \psi}{\partial
r}+\acute{R}\frac{\partial\psi}{\partial
R}-\acute{R}\frac{\partial\tau}{\partial
r}+\acute{A}\frac{\partial\psi}{\partial
A}-{\acute{R}}^2\frac{\partial\tau}{\partial R}-
\acute{R}\acute{A}\frac{\partial\tau}{\partial A},\\
\acute{\phi}=\frac{\partial\phi}{\partial
r}+\acute{R}\frac{\partial\phi}{\partial
R}+\acute{A}\frac{\partial\phi}{\partial
A}-\acute{A}\frac{\partial\tau}{\partial
r}-{\acute{A}}^2\frac{\partial\tau}{\partial
A}-\acute{R}\acute{A}\frac{\partial\tau}{\partial R}.
\end{eqnarray}
In Eq.($\ref{(22T)}$), $B$ is called the gauge function with $D$
defined as
\begin{equation}\label{(222T)}
D\equiv \frac{\partial}{\partial
t}+\dot{a}\frac{\partial}{\partial
a}+\dot{R}\frac{\partial}{\partial R}.
\end{equation}
Substituting ($\ref{z34}$) in Eq.($\ref{(22T)}$) and after some
manipulations, we get an over determined system of linear PDEs, i.e.
\begin{eqnarray}
\label{(DD1)}\tau_A&=&0,\\\label{(DD2)}\tau_R&=&0,\\\label{(DD3)}\psi_A&=&0,
\\\label{(DD4)}\phi_R&=&0,\\
\label{(DDs5)}\phi_r&=&B_R,\\\label{(DD6)} 2r(\tau
f''+f'\tau_r)+r^2(\psi
f'''+\phi_{A}f''+\psi_Rf''-\tau_{r}f'')&=&0,\\ \label{(DD7)} -2\tau
f'-2r(\psi f''+f'\phi_A)+r^2f''\psi_r&=&B_A,\\\nonumber
2r\tau(f-Rf')-2f'\phi-r^2Rf''\psi+2f''\psi-2f''\psi
A-\\2rf'\phi_r+r^2\tau_rf-r^2\tau_rRf'+2\tau_rf'-2\tau_rAf'&=&B_r\label{(DD8)}.
\end{eqnarray}
When $f(R)$ is arbitrary, we obtain trivial results, i.e.
\begin{eqnarray}
\tau=0,~~\psi=0, ~~\phi=0,
\end{eqnarray}
However we use a well known form of $f(R)$, i.e.
\begin{equation}\label{(z22T)}
f(R)=f_0R^n,~~~~~n\neq0,1.
\end{equation}
This function has been widely used in different cosmological
context. The determining equations, using Eq.($\ref{(z22T)}$) yields
\begin{equation}\label{(z229T)}
\tau=c_1r,~~~\psi=-3c_1\frac{R}{n},~~~\phi=\frac{c_1}{n}(2An-3A-2n+3),
\end{equation}
where the gauge term is zero in this case. Thus the Noether
symmetry generator in this case is given by
\begin{equation}\label{(z2299T)}
X=r\frac{\partial}{\partial
r}-3\frac{R}{n}\frac{\partial}{\partial
R}+\frac{1}{n}(2An-3A-2n+3)\frac{\partial}{\partial A},
\end{equation}
and the conserved quantity turns out to be
\begin{equation}
I=r(n-1)\bigg[r^2R^n+6(A-1)R^{n-1}+5r\acute{A}R^{n-1}-
(2n-3)(A-1)r\acute{R}R^{n-2}+r^2\acute{A}\acute{R}R^{n-2}\bigg].
\end{equation}

\section{Summary and Conclusion}

The main objective of this paper is to investigate the Noether
symmetries in metric $f(R)$ gravity. FRW and spherically symmetric
spacetimes are considered for this purpose. We present a general
solution of determining equations for FRW universe with gauge term.
In fact, we get four Noether symmetry generators. A non-zero gauge
term is obtained which depend on Ricci scalar $R$ and scale factor
$a$. It would be worthwhile to mention here that solution already
obtained by Jamil et al. \cite{22} is a subcase for the flat
universe with zero gauge term. Moreover, in palatini $f(R)$ gravity,
a non-zero time dependent gauge term is obtained \cite{platini}. For
non-flat universe, we obtain one symmetry generator which is
translation of time coordinate. The first integrals are obtained in
each case. The interesting feature of cosmological model for
$f(R)=f_0R^{3/2}$ is that it gives a negative deceleration parameter
which is consistent with the recent experimental results to justify
the accelerated expansion of universe. It has been shown
\cite{quintessence} that for $a=a_0t^2$ and $f(R)=f_0R^{3/2}$, the
deceleration parameter is $-\frac{1}{2}$ for the flat universe.
\\\\
The spherically symmetric spacetimes yields a set of eight linear
PDEs. These equations are solved for two cases of $f(R)$: First case
involves an arbitrary function of Ricci scalar which gives a trivial
solution. However, we obtain a non-trivial solution in second case
when $f(R)=f_0R^n$. The corresponding conserved quantity is also
obtained in this case. This cosmological model has been used
extensively in the recent literature. In particular, the well known
$f(R)$ model with inverse curvature term, corresponding to $n=-1$,
predicts late time accelerated expansion of the universe
\cite{Vollick}.
\\\\Moreover, the stability conditions for $f(R)$ models
 are $f'(R)>0$ and $f''(R)>0$ \cite{Starobinsky}. It would be
worthwhile to mention here that the model $f(R)=f_0R^{3/2}$
satisfy these conditions for $f_0>0$ and $R>0$. These conditions
are also satisfied by the model $f(R)=f_0R^n$ when $f_0>0$,
$n-1>0$ and $R>0$. \vspace{1.0cm}
\\\\\textbf{Acknowledgement}\\\\ MFS is thankful to National University
of Computer and Emerging Sciences (NUCES) Lahore Campus, for
funding the PhD programme.

\end{document}